\documentclass[10pt,letterpaper,twocolumn]{article}

\usepackage{ol2}
\usepackage[draft]{hyperref}
\usepackage{amsmath}
\usepackage{graphics}

\newcommand{\gt}[0]{g^{(2)}(0)}
\newcommand{\gtt}[0]{g^{(2)}(n)}
\newcommand{\dhd}[0]{$D_H$ }
\newcommand{\da}[0]{$D_A$ }
\newcommand{\db}[0]{$D_B$ }

\begin{document}

\twocolumn[

\title{A microstructured fiber source of photon pairs \\ at widely separated wavelengths}

\author{Joshua~A.~Slater,$^{1,*}$ Jean-Simon~Corbeil,$^2$ St\'ephane~Virally,$^2$ F\'elix~Bussi\`eres,$^{2,1}$ Alexandre Kudlinski,$^3$ G\'eraud~Bouwmans,$^3$ Suzanne~Lacroix,$^2$ Nicolas~Godbout,$^2$ and Wolfgang~Tittel$^1$}

\address{
$^1$ Institute~for~Quantum~Information~Science~(IQIS), Department~of~Physics~and~Astronomy, University~of~Calgary, $2500$~University~Dr~NW, Calgary, Alberta T$2$N~$1$N$4$, Canada. \\
$^2$ Centre~d'Optique,~Photonique~et~Laser~(COPL), \'Ecole~Polytechnique~de~Montr\'eal, P.O.~Box~$6079$, Station~Centre-Ville, Montr\'eal, Qu\'ebec H$3$C~$3$A$7$, Canada \\
$^3$
IRCICA, FR~CNRS~3024, Laboratoire~de~Physique~des~Lasers, Atomes~et~Mol\'ecules~(PhLAM), Universit\'e~Lille~1, UMR~CNRS~8523, 59655~Villeneuve~d'Ascq~Cedex, France
\\
$^*$Corresponding author: jslater@qis.ucalgary.ca \\
}

\begin{abstract}
We demonstrate a source of photon pairs with widely separated wavelengths, $810$~nm and $1548$~nm, generated through spontaneous four-wave mixing in a microstructured fiber. The second-order auto-correlation function $\gt$ was measured to confirm the non-classical nature of a heralded single photon source constructed from the fiber. The microstructured fiber presented herein has the interesting property of generating photon pairs with wavelengths suitable for a quantum repeater able to link free-space channels with fiber channels, as well as for a high quality telecommunication wavelength heralded single photon source. It also has the advantage of potentially low-loss coupling into standard optical fiber. These reasons make this photon pair source particularly interesting for long distance quantum communication.
\end{abstract}


 ]

Sources of photon pairs constitute a fundamental building block for most emerging quantum communication technologies. They are essential for the development of photonic quantum entanglement~\cite{TW01}, which is required for most quantum communication tasks including Quantum Key Distribution (QKD)~\cite{BB84,Ekert91}, as well as quantum repeaters required to break the distance barrier of QKD~\cite{BDCZ98}
. High quality sources of photon pairs are also needed for Heralded Single Photon Sources (HSPSs)~\cite{HM86}.

Photon pairs have been generated through various techniques including Spontaneous Parametric Downconversion (SPDC) in $\chi^{(2)}$ nonlinear crystals~\cite{BW70}, atomic ensembles~\cite{K03} and in quantum dots~\cite{SYQCRS06}. One drawback to these techniques is the limited coupling efficiency into optical fiber, which impacts on the performance of quantum communication schemes. To avoid this drawback, the generation of photon pairs directly inside optical fiber, through $\chi^{(3)}$ Spontaneous Four-Wave-Mixing (SFWM), has been studied~\cite{FVSK02,kumar05,rarity05,wadsworth09,LCLLVkumar06,DBN09,Migdall08}.

Photon pairs at telecommunication wavelengths ($1550$~nm) have been generated in standard telecommunication Dispersion Shifted Fiber (DSF) by pumping in the anomalous dispersion regime~\cite{kumar05}. In these experiments, the wavelengths of the generated photon pairs are close to the pump wavelength. More recently, photon pairs have been generated with visible and near-infrared wavelengths by pumping in the normal dispersion regime of novel Microstructured Fibers (MSFs)~\cite{rarity05}. A seldom considered source is one that produces photon pairs with one photon in the visible and one photon at telecommunication wavelengths~\cite{wadsworth09,gisin04}. Such a source is necessary to create entanglement for quantum repeaters that can link free-space networks, which generally rely on visible photons, with telecommunication networks, which depend on wavelengths near $1550$~nm. In addition, a high-quality HSPS at telecommunication wavelength can be obtained with a heralding photon at $810$~nm wavelength as high-efficiency detectors are available~\cite{BSGT08}. Also, as the wavelengths of both generated photons lie outside the spectral range of spontaneous Raman scattering of the pump photons, cooling the MSF to cryogenic temperatures, as in previous experiments~\cite{LCLLVkumar06,DBN09}, is not required.

\begin{figure}[!b]
\centerline{\includegraphics[width=8.5cm]{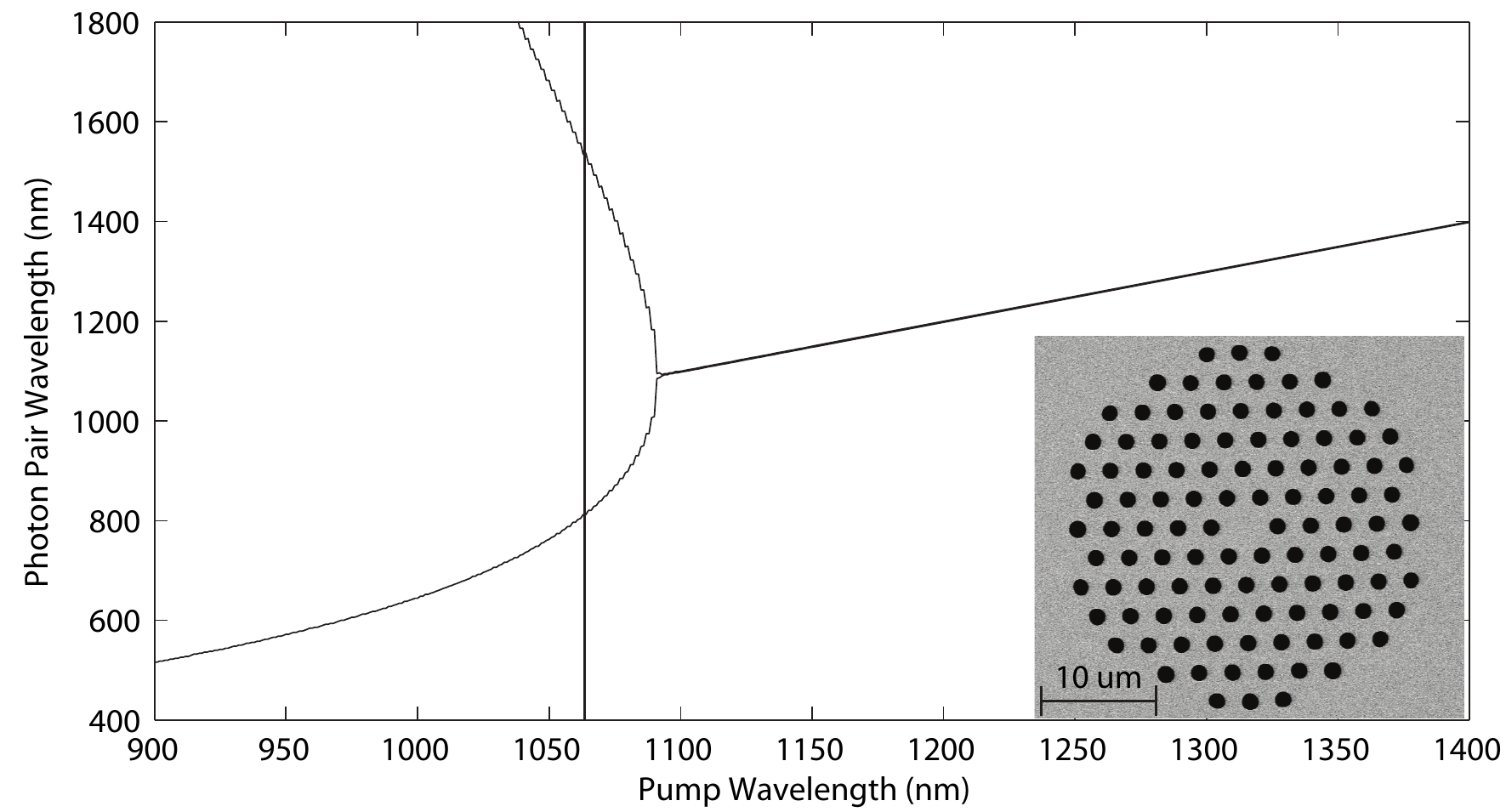}}
\caption{Phase-matching curves of the MSF~\cite{KPBM08}. The vertical line corresponds to the pump wavelength. Inset: Scanning Electron Microscope image of MSF.}\label{fig:phasematch}
\end{figure}
In this Letter we report on the generation of photon pairs with widely separated wavelengths with an MSF. We pumped an MSF, designed with a zero-dispersion wavelength of $1092$~nm~\cite{KPBM08}, with $1064$~nm light and produced photons centered at $810$~nm and $1548$~nm (see Fig~\ref{fig:phasematch}). We confirmed the non-classical nature of the source by measuring classically forbidden values of the second-order auto-correlation function $\gt$ of the $1548$~nm photon field heralded by an $810$~nm photon detection~\cite{Migdall08}. This important measure can be used to assess the visibility of two-photon interference~\cite{SRMZG05,BSGT08} required for quantum repeaters as well as the security of QKD using an HSPS~\cite{WSY02}.

\begin{figure}[!b]
\centerline{\includegraphics[width=8.5cm]{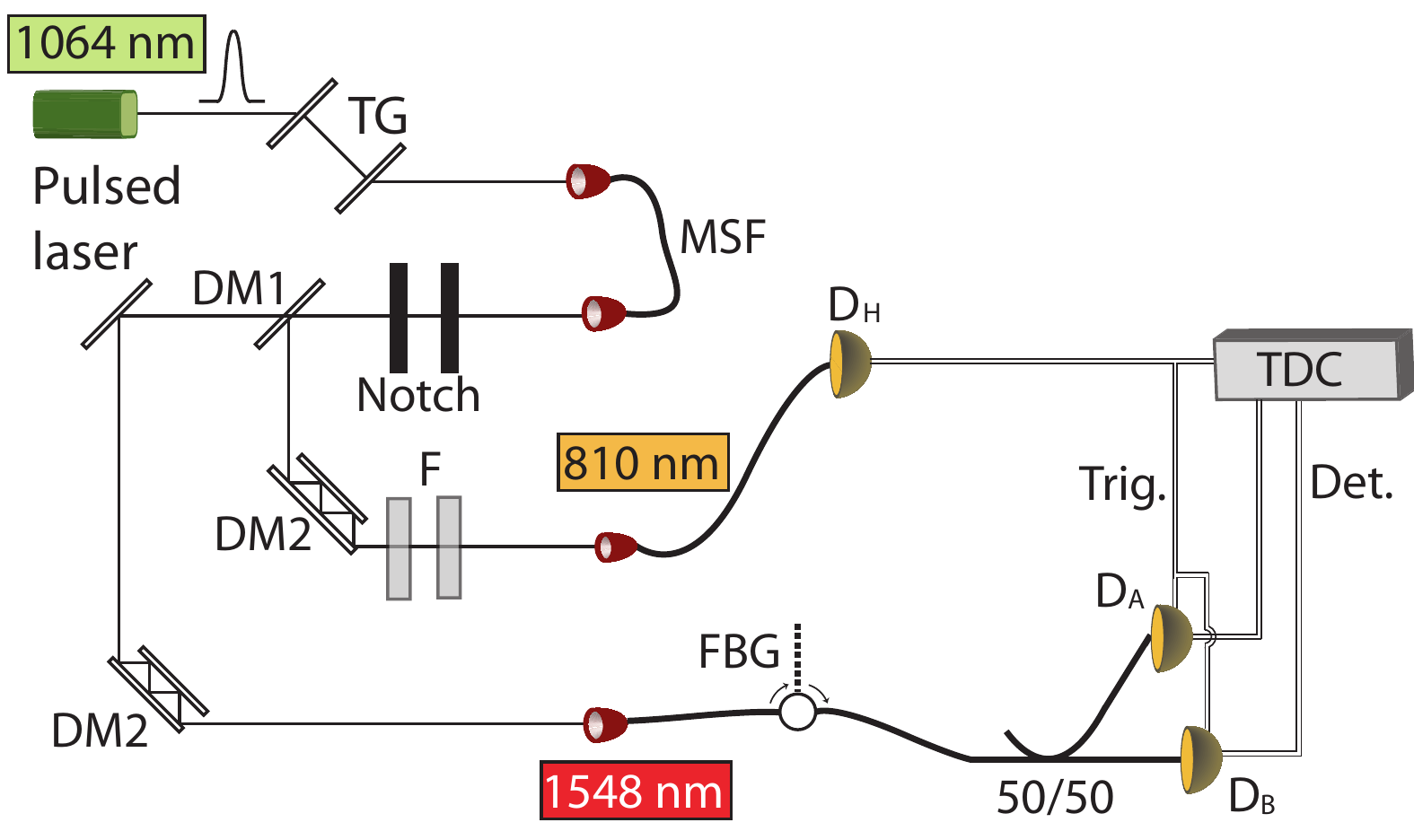}}
\caption{Experimental setup. MSF, microstructured fiber; DM1, dichroic mirror to separate $810$~nm and $1548$~nm photons; $D_H$, silicon-based single photon avalanche photodiode (APD); $D_A$, $D_B$, InGaAs-based single photon APD; 50/50, fiber $50/50$ coupler; Trig., \dhd detection signal used to activate $D_A$ and $D_B$; Det., \da or \db detection signal; TDC, time-to-digital converter used to record all detections; TG, transmission Grating used to purify $1064$~nm pump; Notch, notch filter to reject pump; DM2, dichroic mirror used in multipass configuration to separate $810$~nm or $1548$~nm photons from the pump; F, short-pass and long-pass filter to isolate $810$~nm photons; FBG, fiber Bragg grating to isolate $1548$~nm photons.}\label{fig:expsetup}
\end{figure}

In our experiment (see Fig.~\ref{fig:expsetup}) a mode-locked Nd:YAG laser pumped the $10$~m MSF. The laser created $300$-ps-long pulses at $1064$~nm wavelength with a $100$~MHz repetition rate and was spectrally purified by two transmission gratings. The wavelengths of the generated photons, $810$~nm and $1548$~nm, were verified using an optical spectrum analyzer. The photons were separated using a custom made dichroic mirror. Post-MSF filtering of the pump was accomplished with notch filters, multi-pass dichroic mirrors, edgepass filters and a fiber Bragg grating. This resulted in $>210$~dB isolation of the pump beam and only $7$~dB and $6$~dB loss to the collection fibers for the $810$~nm and $1548$~nm photons, resp. This included $40\%$ coupling efficiency into each fiber. This could be improved by splicing the MSF directly to a standard fiber, where transmissions up to $95\%$ with MSFs similar to ours have been reported~\cite{wadsworth09}, and doing the filtering in fiber. The $810$~nm photons were detected by a Si based single-photon detector ($D_H$) while the $1548$~nm photons were sent to single-photon detectors configured in a Hanbury Brown \& Twiss (HBT) experimental setup~\cite{HBT56}: two InGaAs based single-photon detectors ($D_A$ and $D_B$), which were activated only when \dhd detected an $810$~nm photon (i.e. the $810$~nm photon heralded the $1548$~nm photon), preceded by a fiber $50/50$ coupler. Detection signals were collected by a Time-to-Digital Converter (TDC) only when \dhd detected a photon and \da and \db were not within dead-time. The overall efficiency was $8\%$ and $0.25\%$ for $810$~nm and $1548$~nm respectively. Note that the $10\%$ detection efficiency of \da and \db was a significant contribution to this loss.

\begin{figure}[!t]
\centerline{\includegraphics[width=8.5cm]{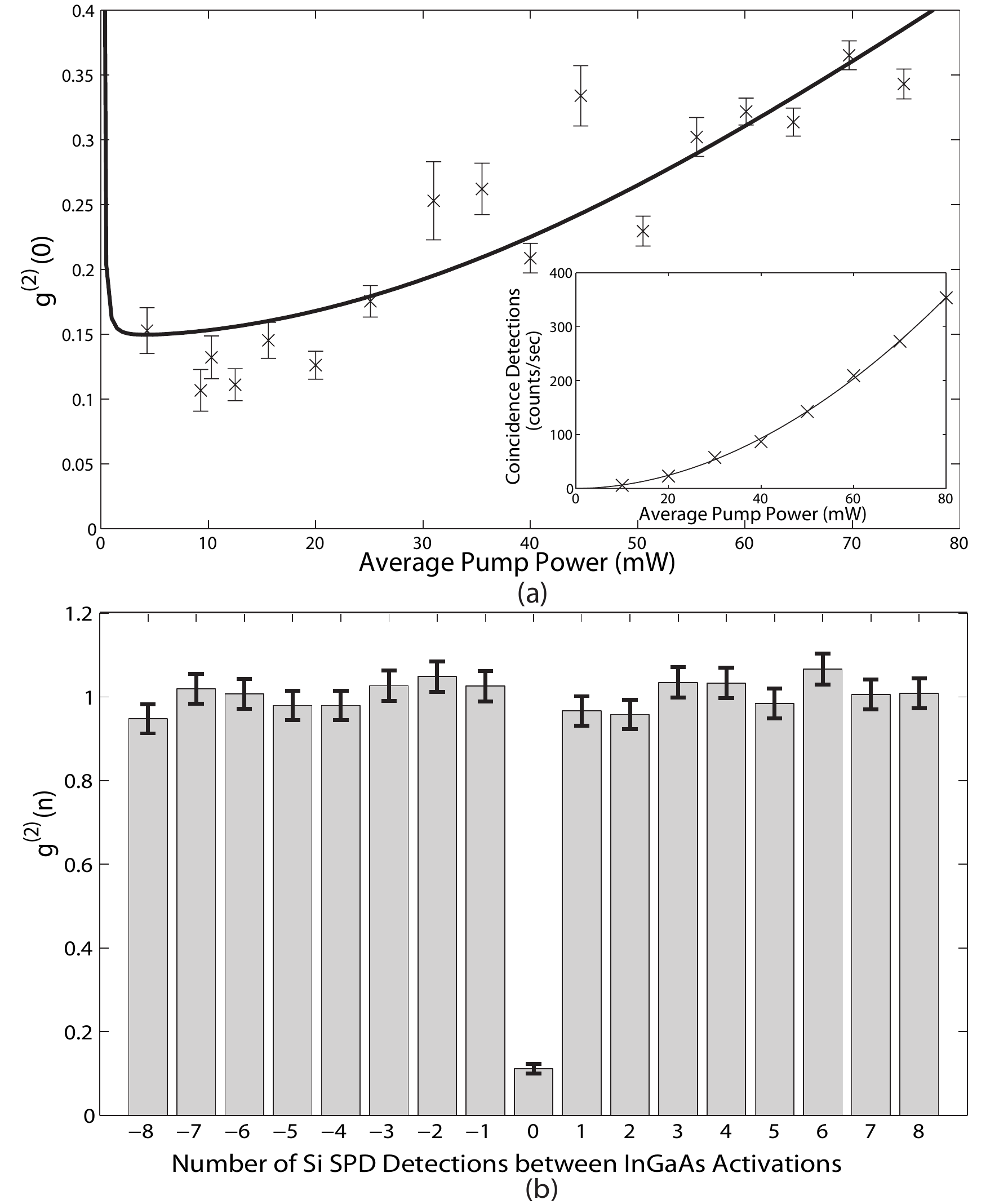}}
\caption{(a) $\gt$ as a function of pump power. Error bars depict one standard deviation. Solid line represents theoretical $\gt$ prediction. Inset: \dhd and \da coincidence detections as a function of power. The quadratic dependency is a characteristic of FWM when $p_{\geq2} \ll p_1$. This data was obtained from a different, yet typical, experimental run than the $\gt$ data. (b) $\gtt$ at $12.5$~mW average pump power ($3.5\times 10^{4}$~$810$~nm counts/sec resulting in $4.4\times 10^{-3}$ photon pairs per pulse), where $n$ is the $D_H$ detection offset between \da and \db activations.}\label{fig:gtwo}
\end{figure}

First, we measured count rates on \dhd and coincidence count rates between \dhd and \da as a function of power. Both have a nearly perfect quadratic dependency, which is a signature of SFWM when the probability of producing multiple pairs is much less than the probability of producing one pair, $p_{\geq2} \ll p_1$ (see Fig.~\ref{fig:gtwo} inset), although this does not demonstrate the nonclassical nature of the source.

We also examined the coincidence detection rate between \dhd and \da as a function of the time delay between a detection at \dhd and the activation time for $D_A$. At $20$~mW average power (approximately $670$~mW peak power), the maximum coincidence rate was $268$~counts/s. Displacing the activation time for \da by integer multiples of the time between subsequent pump pulses produced an accidental coincidence detection rate of $14.6$~counts/s (coincidence detection rates at non-integer multiples were negligible), resulting in a Coincidence-to-Accidental Ratio (CAR) of $268/14.6 \approx 18.3$ (In general, the CAR was between $20$ and $10$). Accidental coincidences were likely due to fluorescence from various optical elements.

To demonstrate the non-classical nature of the source, we measured the second-order auto-correlation coefficient at zero time-delay, $\gt$, of the heralded photon field as a function of pump power. The $\gt$ was calculated according to
\begin{equation}
  \gt = \frac{p_{AB | H}}{p_{A | H} \times p_{B | H}},\label{eq:g2}
\end{equation}
\noindent where $p_{AB | H}$ is the probability for a \da and \db coincidence detection, per pump pulse, provided there was a detection at $D_H$, and similarly for $p_{A | H}$ and $p_{B | H}$~\cite{GRA86,BSGT08}.

Classical theory demands that $\gt \geq 1$; uncorrelated detections yielding $\gt = 1$ and photon bunching from a thermal source yielding \mbox{$\gt = 2$}~\cite{GKqo}. The quantum description of light allows for $0 \leq \gt < 1$: this is known as photon anti-bunching. In particular, a source that emits only one photon pair per pulse could produce $\gt = 0$, if used in an heralding HBT experiment. However, this is not achievable with photon pairs generated from SFWM as there always exists a finite probability that the source emits multiple pairs per pump pulse, which raises $\gt$ above zero. Nevertheless, measuring a $\gt < 1$ confirms the nonclassical nature of the source, and consequently the possibility to use the source in quantum communication protocols.

The results of our $\gt$ measurements are presented in Fig.~\ref{fig:gtwo}. The $\gt$ is well below one, and matches well with theoretical predictions, over a wide range of pump powers. The theoretical curve was calculated using a method similar to~\cite{BSGT08}. A Poisson distribution was assumed for the number of photon pairs as the coherence time of the photon pairs (inferred from bandwidth measurements of $1.6$~nm and $5.8$~nm for the $810$~nm and $1548$~nm photons, respectively) was much shorter than the duration of the pump pulse~\cite{RSMATZG04}. The scatter in the points is mainly due to temperature instabilities in the laser cavity. These variations change the peak power of the laser, which affects the probability to produce a photon pair and thus the $\gt$. These results demonstrate the nonclassical nature of the source along with its suitability for quantum communication.

Fig.~\ref{fig:gtwo}(b) depicts $\gtt$ at $12.5$~mW average pump power for various $D_H$ detection offsets ($n$) between the activation of $D_A$ and the activation of $D_B$ (i.e. when $D_H$ detected a photon $D_A$ was activated such that it could detect the simultaneously created photon, while $D_B$ was activated after $n$ more $D_H$ detections). One observes that $\gtt \simeq 1$ for all $n$ except $n = 0$. This is expected as photons produced by different laser pulses are uncorrelated. The sharp $\gtt$ decrease at $n = 0$ [$\gt = 0.11 \pm 0.01$, with $p_{A | H} = (2.975 \pm 0.006)\times 10^{-3}$, $p_{B | H} = (3.162 \pm 0.006)\times 10^{-3}$ and $p_{AB | H} = (1.05 \pm 0.11)\times 10^{-6}$] is a demonstration of photon antibunching.

In conclusion, we have demonstrated an MSF source of photon pairs with widely separated wavelengths through measurements of the second-order auto-correlation function, which confirmed the nonclassical nature of the source. The source is compatible with existing telecommunication infrastructure and, moreover, can be developed into a source of quantum entanglement suitable for future experiments in quantum communication including quantum repeaters that can link free-space and fiber based quantum channels into one coherent network.

\section*{Acknowledgements}

The authors thank Mika\"el~Leduc and Vladimir Kiselyov for their technical support, Yasaman Soudagar for many useful discussions and NSERC, iCORE, GDC, FEDER, CFI, AET, QuantumWorks, AIF, NATEQ, and CIPI.

\end{document}